\documentclass[newstyle,twocolumn,proceedings]{rmaa}

\usepackage{rmaacite}

\newcommand{\etal}{et~al.}

\newcommand{\lya}{\mbox{Ly$\alpha$}}

\def\msun{\ifmmode {\rm M_\odot} \else M$_\odot$\fi}

\def\msunyr{\ifmmode {\rm M_\odot~yr^{-1}}\else${\rm M_\odot~yr^{-1}}$\fi}
\def\lam{\ifmmode {\lambda} \else {$\lambda$} \fi}
\def\muobs{\ifmmode {\mu_{o}} \else  $\mu_{o}$ \fi}
\def\mdoto{\ifmmode {\dot{M}_0} \else  $\dot{M}_0$ \fi}
\def\teff{\ifmmode {T_{eff}} \else $T_{eff}$ \fi}
\def\ilam{\ifmmode {I_\lambda} \else  $I_\lambda$ \fi}
\def\inu{\ifmmode {I_\nu} \else  $I_\nu$ \fi}
\def\fnu{\ifmmode {F_\nu} \else  $F_\nu$ \fi}
\def\tauh{\ifmmode {\tau_{\rm H}} \else $\tau_{\rm H}$ \fi}
\def\cm{\ifmmode {\rm cm} \else  cm \fi}
\def\cmmitwo{\ifmmode \rm cm^{-2} \else $\rm cm^{-2}$\fi}
\def\cmmithree{\ifmmode \rm cm^{-3} \else $\rm cm^{-3}$\fi}
\def\cmps{\ifmmode \rm cm~s^{-1}\else $\rm cm~s^{-1}$\fi}
\def\cmpsps{\ifmmode \rm cm~s^{-2}\else $\rm cm~s^{-2}$\fi}
\def\kmps{\ifmmode \rm km~s^{-1}\else $\rm km~s^{-1}$\fi}
\def\kmpspmpc{\ifmmode \rm km~s^{-1}~Mpc^{-1} \else
    $\rm km~s^{-1}~Mpc^{-1}$\fi}
\def\ergps{\ifmmode \rm erg~s^{-1} \else $\rm erg~s^{-1}$ \fi}
\def\ergpspcm{\ifmmode \rm erg~s^{-1}~cm^{-2}
  \else $\rm erg~s^{-1}~cm^{-2}$ \fi}
\def\ergpspcmphz{\ifmmode \rm erg~s^{-1}~cm^{-2}~Hz^{-1} \else $\rm
   erg~s^{-1}~cm^{-2}~Hz^{-1}$ \fi}
\def\ergpspcmpa{\ifmmode \rm erg~s^{-1}~cm^{-2}~\AA^{-1} \else $\rm
    erg~s^{-1}~cm^{-2}~\AA^{-1}$ \fi}
\def\ergpsphz{\ifmmode \rm erg s^{-1} Hz^{-1} \else
   $\rm erg s^{-1} Hz^{-1}$ \fi}
\def\eg{e.g.}

\title{Chemical Evolution of Galaxies}

\author{Gregory A. Shields
  \affil{Department of Astronomy, University of Texas at Austin} }

\fulladdresses{
\item G. A. Shields: Department of Astronomy,
  University of Texas,
  Austin, Texas 78712 (shields@astro.as.utexas.edu).}

\shortauthor{G. A. Shields}
\shorttitle{Chemical Evolution of Galaxies}

\keywords{Galaxies:evolution ---  Galaxies:abundances
--- ISM:abundances --- Stars:Abundances}

\abstract{%
Chemical abundances provide important clues to the evolution of 
galaxies.  Ionized nebulae are one of the main sources of chemical 
abundance measurements, especially in external galaxies.  Studies 
of H II regions have shown that the overall metallicity of 
galaxies increases with galactic luminosity, and that spiral 
galaxies characteristically have radial gradients in chemical 
composition.  There are indications of environmental influences 
on chemical abundances.  Planetary nebulae provide another measure 
of abundances in the Milky Way and other galaxies.  Space 
facilities have allowed measurements for elements that 
are inaccessible at optical wavelengths.  Large telescopes 
make possible the study of individual stars in external galaxies 
and the study of interstellar abundances in galaxies at 
intermediate and high redshifts.  These advances promise exciting 
times as astrophysicists strive to paint a complete picture of 
galactic evolution from the Big Bang to the present.
 }

\listofauthors{G.~A.~Shields}
\indexauthor{G.~A.~Shields}

\begin{document}

\maketitle

\section{Introduction}
\label{sec:intro}
Ionized nebulae have played an important role in the
measurement of interstellar abundances in our own galaxy and others while also
serving as a laboratories for atomic physics.
Early nebular studies focused on planetary nebulae and H II regions in our
galaxy.  These showed abundances roughly similar to those in the sun, but with
significant differences depending on the element, location in the Galaxy, and
the stellar population involved.  The advent of sensitive detectors and the
increasing availability of large telescopes made possible the systematic study
of giant extragalactic H II regions (GEHRs), planetary nebulae, and
supernova remnants in external galaxies.  This work revealed patterns of
variation of chemical composition with position in a galaxy, and from galaxy
to galaxy.  Today we know that chemical abundances decrease outward across
the disks of galaxies, and that they increase with increasing galactic
luminosity.  These trends are qualitatively echoed in the abundances of stars
in elliptical galaxies.  The relative abundances of the elements also show
systematic trends.  There are indications that the cluster environment
affects the chemical evolution of spiral galaxies.  Observations of QSO
emission lines suggest a metal rich environment in galactic nuclei even at
high redshift.  Studies of QSO absorption lines probe the interstellar medium
of galaxies at early times.  These results provide a rich set of
constraints for theoretical models.
For a 
detailed discussion of abundances in spiral and elliptical galaxies, the
reader is referred to the excellent review by Henry \& Worthey (1999,
``HW'').  Observational results and theoretical foundations are discussed in
the classic review by Tinsley (1980). 

\section{Foundations}
\label{sec:foundations}
 
A useful reference model for chemical evolution theory is
the so-called ``simple
model'', involving progressive conversion of gas to stars in a
closed box.  In this model, stars either live
forever or die instantaneously, returning enriched material to the
interstellar medium (ISM).  The mass of freshly produced metals per unit mass
of long lived stars formed is the ``yield'', $p$.  Then the mass fraction
of metals, $Z$, in the gas increases as the gas fraction, $\mu \equiv
M_{\rm gas}/M_{\rm tot}$, decreases, according to $Z = p\, \rm ln\, \mu^{-1}$ 
(Searle \& Sargent 1972). An interesting way to express
measured abundances is the ``effective yield'', defined
in terms of the simple model of chemical evolution: $Z(O) = p_{\rm
eff}\rm(O)\, ln\,\mu^{-1}$, etc.

Oxygen, neon, and sulfur, among other elements, are ``primary'' products of
stellar nucleosynthesis, whose yields depend relatively little on the
abundances in the progenitor star.  On the other hand, nitrogen largely
results from secondary production, so that the yield is proportional to the
abundance of C and O in the progenitor.  This leads to an increase in N/O
with increasing O/H, above a minimum N/O that corresponds to the primary
contribution to nitrogen production.  This is shown by observations of H II
regions in spiral and irregular galaxies for O/H above a threshold value
$\sim0.2$ solar (\eg, Talbot \& Arnett 1974; Garnett 1990, 2001b). Iron is
believed to be produced in part by type Ia supernovae that require time
($\sim 10^9$ yr) to evolve to the point of explosion (see review by Wheeler,
Sneden, \& Truran 1989).  The more massive stars producing elements such as O
explode effectively instantaneously, and therefore the O/Fe ratio is an
indicator of the timescale on which a population of stars was formed.

The number of
main sequence stars as a function of abundance in the solar neighborhood
disagrees with the simple model in the sense that there are too few metal
poor stars (van den Bergh 1962; Audouze \& Tinsley 1976).  This
``G dwarf problem'' is an important constraint on models for chemical
evolution of the Galaxy.  One solution is that continuing infall of metal
poor gas causes the abundances to approach an asymptotic value at which most
stars are formed (Larson 1972).

Pagel (2001) gives a succinct discussion of results involving
stellar abundances in the halo, ``thick disk'', and ``thin disk'' of the
Milky Way.  The high O/Fe ratio in the halo and thick disk suggests a brief
formation period, early in the history of the Galaxy.  The thin disk has
lower O/Fe and a metallicity distribution consistent with prompt initial
enrichment (perhaps from the thick disk) combined with infall of
extragalactic gas, with a hiatus in star formation between the thick and thin
disk formation of several billion years.  See Pagel (2001) and Chiappini,
Matteucci, \& Gratton (1997) for further discussion and references.

\section{Radial Gradients}
\label{sec:gradients}

Radial gradients in chemical composition are characteristic of spiral
galaxies.   Searle (1971) studied the systematic
variation of the [N~II], [O~II], and [O~III] emission-line intensities 
of GEHRs in
several spiral galaxies as a function of galactocentric distance.  His
analysis indicated radial decreases in O/H and 
N/O.  The latter was consistent with the suggestion by Peimbert (1968) that
the strong [N~II] emission from the nuclei of M51 and M81 reflected a high
abundance of nitrogen.  Searle's work was followed by other
observational studies and by analyses involving computer models of the
nebular ionization and thermal structure (\eg, Smith 1975; 
Shields \& Searle 1978).  As results for a substantial number of galaxies
became available (e.g., McCall, Rybski,
\& Shields 1985; Zaritsky, Kennicutt,
\& Huchra 1994, ZKH), analysis of the systematics of abundances
became possible.     Radial gradients may be fit by an
exponential in R/R$_0$, where R$_0$ is the isophotal radius. Barred spirals
have gradients substantially shallower than normal spirals, but similar
overall abundances (Martin
\& Roy 1994; ZKH).  Low surface brightness spirals have relatively low
abundances for their mass (McGaugh 1994).  Local abundances increase with the
local surface surface brightness or surface mass density (McCall
1982; Edmunds \& Pagel 1984; Vila-Costas
\& Edmunds 1992).   

The Milky Way has a gradient in O/H of -0.06 dex/kpc (HW, and references
therein).  This is supported by studies of H II regions and planetary
nebulae (Maciel \& K\"oppen 1994).  Carraro, Ng,
\& Portinari (1998)  note that a variety of theoretical models for the
chemical evolution of the Galaxy match the radial gradient at the present
time, but have quite different predictions for the time evolution of the
gradient.
 They conclude from the
available stellar data, in particular for open clusters, that the gradient has
changed relatively little with time.  This conclusion seems in harmony with
the fact that non-barred spirals have similar gradients, which might not be
expected if gradients changed greatly during the course of galactic evolution.

Space observatories have made possible the measurement of carbon emission
lines in H II regions.  Results for spiral and irregular galaxies have
revealed an unexpected trend in which C/O increases substantially with
increasing O/H (Garnett \etal\ 1999;
Garnett 2001b; and references therein).  This may be explained in terms of
the effect of stellar winds on the evolution of massive stars.  The winds
carry off more mass at higher abundances, causing the evolution of the
stellar core to be arrested and the yield of oxygen to be decreased, relative
to that of carbon (\eg, Carigi 1996).

\section{Abundance Fluctuations}

Chemical
enrichment involves supernova explosions, stellar winds, and planetary nebula
ejections.  Events involving massive stars are
concentrated in young star clusters, leading to the potential for localized
enrichment.  Infall of primordial (or not so primordial) gas from the
environment of a galaxy may involve accretion of gas clouds or dwarf
galaxies, possibly leading to localized depressions of
heavy element abundances.  Heavy elements are dispersed by
flows propelled by supernova explosions and stellar winds, as well as
differential rotation, large scale flows induced by bars, etc. 

Local abundance fluctuations in spiral galaxies have
a smaller amplitude than the overall radial gradients.  There is, however,
evidence for significant fluctuations in
the Milky Way.  The solar oxygen abundance [O/H] = 8.87 (Grevesse \& Noels 
1993) significantly exceeds the Orion nebula value 8.58 (Baldwin \etal\
1991; Esteban \etal\ 1998) as well as HW's ``composite'' H II region value
$8.68 \pm 0.05$ at the solar circle.  (One uncertainty is the nagging question
of temperature fluctuations in the nebular gas [Peimbert 1967]).  Abundances
in B stars in the vicinity of the Orion complex agree with the nebular value,
although there is evidence that abundances increased as star formation
progressed through the region (Cunha
\& Lambert 1992).  The sun is 0.17 dex more metal rich than the average
nearby star (Wielen, Fuchs, \& Dettbarn 1996).  Wielen \etal\ propose that the
sun was born $\sim$2 kpc closer to the Galactic center than its current
orbital radius, where abundances were higher.  Dispersions among open
clusters presumably are less affected by this process, if indeed it is
important (Garnett \& Kobulnicky 2000). 

Edvardsson \etal\ (1993) found a large spread in [Fe/H] in G stars at
the solar circle, but little trend in metallicity with age.  Friel \&
Boesgaard (1992) studied C and Fe abundances in F stars in open clusters in
the Galactic disk.  They found no significant dispersion in abundance among
the stars in a single cluster at the level of 0.05 dex but 
significant differences from cluster to cluster at a level of $\sim0.1$~dex. 
They found little dependence on age from 50 million to 5 billion years.   The
open clusters studied by Carraro \etal\ (1998)  show a range in [Fe/H] of 0.6
dex, suggesting a dispersion $\sigma \approx 0.15$.  In contrast,
interstellar oxygen measurements by Meyer, Jura, \& Cardelli (1998) show a
 dispersion of only
$\pm0.05$~dex on a number of sightlines. 
In a recent study of the Galactic H II regions by Deharveng \etal\
(2000), the scatter of the best measurements around the mean gradient in O/H
is small.  Garnett \& Kobulnicky (2000) 
examine evidence for abundance fluctuations in H~II regions, stars, and the
interstellar medium.   They argue that selection effects in the star sample
of Edvardsson \etal\ complicate the interpretation of the abundance scatter,
and that the true disperson for field stars is less than 0.15~dex. 

Results for external spirals are inconclusive, because of the lack of
precise measurements of the electron temperature in a sufficient sample of
GEHRs.  
In a study of H II regions in M101, Kennicutt
\& Garnett (1996) suggest, on the basis of line ratio correlations, that the
true dispersion around the mean gradient is less than the observed scatter of
0.1 to 0.2 dex. 

These results suggest that, in the disks of spiral
galaxies, abundance fluctuations with $\sigma \approx$ 0.1~dex occur on a
scale larger than the clouds that form a typical open cluster, but smaller
than the distance between open clusters.  The dominant cause of the
fluctuations remains unclear.  Several processes have been discussed,
including local enrichment by supernovae and accretion of clouds of metal poor
gas (\eg, Carraro \etal\ 1998, and references therein). 
Franco \etal\ (1988) discuss the possibility that the Orion
molecular cloud complex, with its relatively low metallicity, results from
the impact of a metal poor gas cloud.  

The prevailing level of abundance fluctuations in the ISM depends on
the competition between localized enrichment and mixing.   Roy
\& Kunth (1995)  considered mixing on different scales in the Galactic disk. 
On scales of 1 to 10 kpc, local abundance fluctuations are smoothed in less
that $10^9$~yr by turbulent diffusion of clouds together with differential
rotation.  On scales 100 to 1000 pc, cloud collisions, star-formation driven
flows, and differential rotation mix the gas in $10^8$~yr.  On scales 1 to
100 pc, turbulence in ionized regions mixes the gas in $10^{6.3}$~yr or less.

Abundances in irregular galaxies generally are quite uniform. 
Kobulnicky
\& Skillman's (1997) results for NGC 1569 show a dispersion of $\pm
0.05$~dex in O/H and N/O in the well measured regions. They argue that, at
the low metallicity of this object, pollution by the ejecta of only a few
massive stars should give measurable enrichment.  They suggest that the
supernova ejecta form a hot wind that escapes the galaxy, disperses the
ejecta, and later reaccretes, avoiding local enrichment at the site of the
stars' deaths.  There are, however, rare cases of local
enrichment, in particular an H II region with high N/O in NGC 5253
(Kobulnicky \etal\ 1997).  

\section{Luminosity Dependence}
\label{sec:luminosity}

Characteristic abundances
increase with galactic luminosity,
and this trend embraces irregulars as well as spirals (Skillman \etal\
1989; Garnett 1999; ZKH).  This trend may involve the
escape of nucleosynthesis products in winds from galaxies with low escape
velocities (Matteucci \& Chiosi 1983).   Garnett (2001a) plots
the dependence of effective yield on luminosity;
effective yields for low surface brightness galaxies (LSBs) as a class fall
below those for normal galaxies.  Stellar abundances in elliptical galaxies
also increase with galactic luminosity (HW, and references
therein).

\section{Effect of Environment}
\label{sec:environment}

Modern ideas about the evolution of galaxies emphasize the role of accretion
and mergers.  What is the influence of environment on the chemical
evolution of galaxies?  Skillman \etal\ (1996) studied
the abundances in spirals in the Virgo cluster, based on spectra of GEHRs. 
They found that galaxies in the cluster core, with marked H I deficiencies,
were more metal rich than spirals with normal H I content located in the
periphery of the cluster and in the field.  They suggested that this
results from the curtailment of metal-poor infall onto galaxies in the
cluster core, where they are immersed in the hot cluster medium.  In
contrast, infall continues onto spirals in the cluster periphery and in the
field, restraining the increase in abundances with time.

The question of infall onto the Milky Way relates to the
proposal by Blitz \etal\ (1999) that some of the Galactic high velocity clouds
(HVCs) observed in H~I, and sometimes seen in heavy element absorption lines,
are actually a population of dwarf galaxies or gas clouds belonging to the
Local Group.  The inferred accretion rate of these clouds onto the Milky Way
is interesting in the context of chemical evolution.  The Blitz model
draws together a number of aspects, including cosmological predictions of
dwarf galaxy counts and the dynamical evolution of the Local Group.  However,
the massive H I clouds involved in this model
have not been found in other groups of galaxies (\eg, Zabludoff 2001;
Zwaan 2001).

Also of interest is chemical evolution in low density environments.  Peimbert
\& Torres-Peimbert (1992) obtained spectra of emission-line galaxies in
the Bo\"otes void.  Most appear to be irregular galaxies with H II regions
ionized by OB stars.  Several of the objects have rather low N/O values,
compared to a small sample of nearby H II regions with similar O/H,
including the LMC.  As noted by Garnett (1990), a low N/O value
may occur in a young galaxy or in one whose oxygen has been enriched by a
recent starburst, because N comes largely from lower mass stars with longer
lifetimes than stars producing O.  Thus  Peimbert
\& Torres-Peimbert suggest that the Bo\"otes objects
with low N/O may be young galaxies; and they note that late collapse of
density clumps to form galaxies may be a natural occurrence in a low density
environment.  However, the range of N/O values shown by Garnett (1990)
encompasses the Bo\"otes values, albeit for slightly lower O/H. 

The study of
abundances in galaxies in low and high density environments deserves more
attention.  These studies will require spectra of adequate sensitivity and
wavelength resolution together with measurements of the gross properties of
the galaxies, so that comparisons can be made in a way that isolates the
effect of environment.

\section{Abundances in the Early Universe}
\label{sec:early}

Abundances at high redshift can now be measured in a variety of ways
involving emission and absorption lines, thanks to large telescopes and
sensitive light detectors.  Once the high redshifts
of quasars were recognized, the potential to use these
remarkable objects as probes of the early universe was clear.  Derivation of
abundances from the emission lines of AGN is difficult, however. The width of
the broad lines impedes measurement of weak lines, and the high electron
density in the broad line region (BLR) and the large optical depths of some
of the lines makes analysis difficult.  This is true in particular for the
measurement of the electron temperature, necessary to calculate the line
emissivity.  Shields (1976) noted that relative abundances of C, N,
and O could be derived with less sensitivity to the uncertainty in the
electron temperature.  The N/O and N/C ratio might in turn be an indicator of
the overall metallicity, given the secondary nature of nitrogen production. He
found high N/C in two QSOs and suggested a parallel to high nitrogen
abundances in nearby normal galactic nuclei and AGN.  
Hamann and Ferland (1993) studied QSO abundances as a function of redshift by
bringing together chemical evolution models and photoionization models.  They
concluded that most luminous, high redshift QSOs ($z \approx 2$ to 4) have
abundances higher than solar.  Hamann \& Ferland also noted that iron
abundances in QSOs at high redshift might constrain cosmological models.  

Abundances are also measured in absorbing gas clouds on the line of sight to
a QSO.  This includes the high column density ``damped \lya'' systems (DLAs)
and the \lya\ forest.   An example of the study of heavy element abundances
in DLAs is the work of Pettini \etal\ (1999).  They find [Zn/H]
$\approx$ -1.2 in the redshift range z = 1 to 3, with rather little systematic
dependence on redshift.  The relative abundances of the heavy elements are
consistent with a mild degree of depletion onto grains, and [Si/Zn] is
essentially solar. (The ratio Si/Zn is a useful surrogate for O/Fe.) Thus,
these absorbers do not appear to share the [O/Fe] enhancement of metal poor
stars in the Galactic halo.  Evidently, even at redshifts of 2 or so,
past star formation in the DLA systems had proceeded gradually enough that
iron production kept pace with oxygen and other SN II products.

\lya\ forest clouds
often show C IV $\lambda\lambda 1548, 1550$ absorption lines when
observed with sufficient resolution.  Songaila \& Cowie (1996) measured O VI
as well and found that ionization ratios and line widths pointed to
photoionization rather than collisional ionization.  Mean abundances ratios
appear to be [C/H] $\approx$ -1.5 and [O/C] $\approx$ [Si/C] $\approx$ +0.5,
with a spread of order a factor 3 in C/H (Songaila \& Cowie 1996;
Dav\'e \etal\ 1998; Ellison \etal\ 2000).  These authors note that the high
O/C resembles old halo stars; but as discussed above, O/C is also high in
metal poor H II regions. 

Abundances in DLAs resemble those measured in
GEHRS in the more metal poor dwarf irregulars and in the outermost disks of
spirals, that is, [O/H] $\approx$ -1.5.  Silk, Wyse, \& Shields (1987)
suggested that this widespread minimum may result from reaccretion of enriched
gas lost from an early population of dwarf galaxies.  The more metal poor
stars in the Galactic halo might then represent the cannibalized remains of
the primordial dwarf galaxies.  The very metal poor clouds of the
\lya\ forest must then largely have escaped the reaccretion process, at least
at the epoch at which they are observed.

Observations of abundances in ionized gas in galaxies at high redshifts are
also becoming available.  Kobulnicky \& Zaritsky (1999) observed a sample of
emission-line galaxies at redshifts z = 0.1 to 0.5 and found them to be only
marginally more metal poor than the metallicity-luminosity relation observed
for low redshift galaxies.  In contrast, Kobulnicky \& Koo (2000) obtained
near infrared observations of two \lya\ emitting galaxies at z = 2.3 and 2.9,
finding 12 + log O/H = 8.2 to 8.8.  From these and other data, they find that
emission-line galaxies at these redshifts have abundances substantially below
the present day metallicity-luminosity relation.  They note that these low
abundances resemble those of metal rich globular clusters and raise the
possibility that these objects may resemble the formation of galaxies like the
Milky Way at z
$\approx$ 3.  The emission-line galaxies at high redshift have abundances
higher than the DLAs, indicating that they represent different environments.

\section{Discussion}
\label{sec:discussion}

Progress in understanding the chemical evolution of the Milky Way and other
galaxies has relied on an increasing wealth of
observations and a variety of theoretical inputs.  Modern ideas depart from
the concept of a monolithic collapse of the protogalaxy (Eggen,
Lynden-Bell, and Sandage 1962), embracing
mergers, cannibalism, infall, and radial flows in the disk.
This complicated set of processes must be constrained by a commensurate set
of observations.  For the Milky Way, stellar abundances as a function of age
and kinematics are increasingly available.  Evidence for mergers
in the history of the Galaxy can be found in the kinematics of ``star
streams'' (Majewski 2000).  Abundance measurements of individual stars in
external galaxies are made possible by large telescopes and modern
spectrographs (\eg, Venn \etal\ 2001). 

Ionized nebulae remain the best way to
measure the abundances of many elements in the interstellar gas of galaxies.
With sensitive detectors on large telescopes, this work will increasingly
involve objects at significant redshifts and in a variety of environments,
and precise measurements of local abundance fluctuations on various length
scales.  Ionized nebulae will continue to play an important role in the
description of the chemical evolution of the universe from its youth to the
present.

\acknowledgments

I am indebted to Don Garnett, Chip Kobulnicky, Bernard Pagel, Evan Skillman,
and Chris Sneden for helpful discussions.

\end{document}